\documentclass{ws-ijmpd}
\usepackage[super,compress]{cite}
\usepackage{amssymb}
\begin{document}
\markboth{Victor Berezin, Vyacheslav Dokuchaev and Yury Eroshenko}
{Spherically symmetric double layers in Weyl$+$Einstein gravity}

%
\catchline{}{}{}{}{}
%

\title{SPHERICALLY SYMMETRIC DOUBLE LAYERS \\ IN WEYL$+$EINSTEIN GRAVITY}

\author{VICTOR BEREZIN}
\address{Institute for Nuclear Research of the Russian Academy of Sciences \\
	Prospekt 60-letiya Oktyabrya 7a, Moscow, 117312, Rissia \\ 
    berezin@inr.ac.ru}

\author{VYACHESLAV DOKUCHAEV}
\address{Institute for Nuclear Research of the Russian Academy of Sciences \\
	Prospekt 60-letiya Oktyabrya 7a, Moscow, 117312, Rissia \\ \vskip0.2cm
	National Research Nuclear University ``MEPhI'' (Moscow Engineering Physics Institute) \\
	Kashirskoe shosse 31, Moscow, 115409, Russia  \\
	dokuchaev@inr.ac.ru}

\author{YURY EROSHENKO}
\address{Institute for Nuclear Research of the Russian Academy of Sciences \\
	Prospekt 60-letiya Oktyabrya 7a, Moscow, 117312, Rissia \\ 
	eroshenko@inr.ac.ru}

\maketitle

\begin{history}
\received{27 December 2018}
\revised{Day Month Year}
\end{history}

\begin{abstract}
The role of exact solutions in gravitational theories is impossible to overestimate. But the intrinsically nonlinear character of gravitational equations makes solving them a very difficult and problematic task. This is why the investigations of hypersurfaces where the matter energy-momentum tensor undergoes some discontinuities are so important. The physically interesting discontinuities are jumps (they can be viewed as an idealization of the shock waves) and thin shells (i.\,e., $\delta$-function distributions, describing some idealized matter sources including the potential barriers between two different phases during cosmological phase transitions). In this paper we undertook the thorough investigation of the matching conditions on the singular hypersurfaces in Weyl$+$Einstein gravity. 
Unlike in the General Relativity, where the singular hypersurface may contain, at most, the Dirac $\delta$-function both in the matter energy-momentum tensor in the right hand side of the Einstein equations and in the second derivatives of the metric tensor in their left hand side, in the so called quadratic gravity (the gravitational action integral, except the Einstein-Hilbert and cosmological terms, has all possible quadratic combinations of the Riemann curvature scalar). there appear the possibility of the double layer, i.\,e., the derivative of the $\delta-function$.
But, this derivative is absent in the energy-momentum tensor (no mass-dipole analogous to the charge-dipoles in the classical electrodynamics), so the double layer is a purely geometrical phenomenon ant it may be treated as the purely gravitational shock wave.
The mathematical formalism for the double layers was elaborated layers was elaborated by J.~M.~M~Senovilla for the double layers was elaborated for the generic quadratic gravity. Our choice  of the Weyl$+$Einstein gravity is motivated by thee fact that the latter differs from the generic case in some aspects and requires separate consideration. Moreover, we confined ourselves by the spherically symmetry, because in such a case the theory becomes, essentially, two-dimensional. The three-dimensional hypersurface reduces to the world-line, and it becomes much easier to understand every step in the calculations and interpretations of the results.
The main results are the following. We derived the matching conditions for the spherically symmetric singular hypersurface (in our case it is equivalent to the world line) in the Weyl$+$Einstein gravity. It was found, that the residual extrinsic curvature tensor of this surface (i.\,e., having only one, $(00)$)-component after separating the spherical angular part of the metric tensor), indeed, must be continuous on the singular hypersurface. The result is the same as that found by Senovilla, it is dictated by the very possibility to have the double layer, but the jump in the normal derivative of the radius may be not zero. (and this is different from Senovilla's). 
It was found that in the presence of the double layer, the matching conditions contain an arbitrary function (in our case it is a function of the proper conformal time of the observer, sitting on the singular world-line), and this result is quite new and very important. One of the consequences of such freedom is that the trace of the extrinsic curvature tensor of the singular hypersurface is necessarily equal to zero.
We suggested the physical interpretation for the $S_0^n$ and $S_n^n$ components of the surface matter energy-momentum tensor of the shell. In General Relativity they are zero by virtue of the Einstein equations. In the quadratic gravity they are not necessarily zero. Our interpretation is that these components describe the energy flow $(S^{0n})$ and the momentum transfer $(S^{nn})$ of the particles produced by the double layer itself. Moreover, the requirement of the zero trace of the extrinsic curvature tensor (mentioned above) implies that $S_n^n=0$, and this fact also support our suggestion, because it means that for the observer sitting on the shell, the particles will be seen created by pairs, and the sum of their momentum transfers must zero.
We derived also the matching conditions for the null hypersurface, and this is, again, quite new. We found that the null-double layer in the Weyl$+$ Einstein gravity does not exist at all.
\end{abstract}
\ccode{PACS numbers: 04.20.Fy, 04.20.Jb, 04.50.Kd, 04.60.Bc, 04.70.Bw.}

\tableofcontents

\section{Introduction}	

The necessity of considering the geometry of $d$-dimensional hypersurfaces immersed  into the $(d+1)$-dimensional space-time arises whenever one needs to have some slicing  like in the Hamiltonian (canonical) description of General Relativity. The most striking example is the famous Arnovitt-Deser-Misner (ADM) formalism. Or, when the hypersurface divides the whole space-times into two parts with different structures of the matter energy-momentum tensor. In such a case this hypersurface, $\Sigma$, describes either the propagation of the shock waves, if the components of the energy-momentum tensor $T_{\mu\nu}$ are jumped across $\Sigma$, $[T_{\mu\nu}]|_{\Sigma}\neq0$, or it describes the trajectories of the thin shells, if the energy-momentum tensor has a singularity, i.\,e.,  $[T_{\mu\nu}]|_{\Sigma}=S_{\mu\nu}\delta(\Sigma)$, where $S_{\mu\nu}$ is called the the surface energy-momentum tensor, $\delta(\Sigma)$ is the Dirac delta-function, and $[{\;\,}]\equiv (+)-(-)$ denotes the jump. 

Also, special attention should be paid to the boundary surfaces when the least action principle is used for deriving the equations of motion in the cases, when th Lagrangian of the theory under consideration contains the second and linear derivatives of the fields subject to variations. In what follows we will be dealing with the four-dimensional space-time and, accordingly, with the three-dimensional hypersurfaces. 

Since the energy-momentum tensor may have discontinuities on the hypersurface $\Sigma$, one needs the matching condition on it for the gravitational field equations. For the time-like thin shells in General Relativity the corresponding formalism was elaborated by W. Izrael \cite{Isr,Cruz67} in 1966, and since then it was widelly used for investigations of cosmological phase transitions (see \cite{Lake79,BKT83,BKT83b,BKT83c,BKT84,Gyulassy84,Lake84,Grand84,Legovini84,BKT85,Sato86,Laguna86,BKT87,BKT87b,Blau87,Farhi87,Aurilia89,Farhi90,BKT90,BKT91,BKT91b,Aguirre05,Lee07,Mann08} for more details). and black hole structure (e.\,g., see \cite{Boulware73,Frolov74,Berezin05,Dokuch07,Dokuch08,Chernov08,Dokuch10,Berezin14,Berezin14b}). It is assumed that, by the suitable coordinate transformations on both sides of the matching surface, the metric tensor $g_{\mu\nu}$ can be made continuous across $\Sigma$  --- otherwise we would deal with two disconnected manifolds. Thus the matching surface $\Sigma$ is characterized by its own three-dimensional metric tensor, $\gamma_{ij}$ (the Greek indexes label quantities in the four-dimensional space-time, while the Latin ones refer to that on the three-dimensional matching surface), and by the extrinsic curvature tensor, $K_{ij}$, that shows how this hypersurface is embedded into the four-dimensional space-time. And the Israel equations relate the surface energy-momentum tensor, $S_{ij}$, to the jumps in the extrinsic curvature tensor $[K_{ij}]$ across $\Sigma$.

The Einstein equations are of the second order in derivatives of the metric tensor. Hence, the first derivatives may have, at most, the jumps, while the second derivatives may exhibit, at most, the delta-function behavior on $\Sigma$. Therefore, the most singular of the allowed types of the matching hypersurfaces, containing the nonzero energy-momentum tensor, are just the thin shells. As a consequence the thin shells produce the real space-time singularities where the Riemann curvature tensor diverges. The extension of the Israel's formalism to the space-like case is quite obvious. The space-like thin shells were used for describing the very process of the phase transitions in the cosmology by V. A. Berezin, V. A. Kuzmin and I.~I.~Tkachev in Ref.~\refcite{BKT91b} and by V. P. Frolov, M. A. Markov  and V. F. Mukhanov in Ref.~\refcite{FMM}.

The matching hypersurface can, of course, be null. But, in this case the above formalism is not adequate because it is based on the existence of the unit normal vector across $\Sigma$, and the normal vector to the null surface lies just on this very surface. The matching conditions on the null surfaces were derived by one of the authors in Ref.~\refcite{Berezin} and the covariant approach was developed by C. Barrabes and W.~Israel in Ref.~\refcite{BI}.

Started in 1969, there was a long history and big activity in studying the quantum field theories in curved space-times (for more details see \cite{Parker69,Grib70,Zeld70,Zeld71,Zeld72,Parker73,Fulling73,Parker73b,Parker74,Parker74b,Parker74c,Lukash74,Zeld77} and references therein). Here the most interesting for us is the main result, that the renormalization procedure, inevitable in quantum field theory, leads to the appearance in the effective gravitational action of some additional (to the standard Einstein-Hilbert Lagrangian, linear in curvature scalar) terms quadratic in the Riemann tensor and its contractions (Ricci tensor and curvature scalar). When the gravitational Lagrangian is the general combination of these quadratic terms plus the linear term and plus the cosmological constant, it is usually called ``the quadratic gravity'' . Some of its special form was used by A. A. Starobinsky in Ref.~\refcite{Alstar} for constructing the first inflationary cosmological scenario. 

The necessity of considering the quadratic and higher order terms in the effective gravitational Lagrangian was foreseen by A. D. Sakharov in Ref.~\refcite{Sakh} in 1967. He even suggested that the gravitation itself is just the vacuum tensions and stresses of all other quantum fields. Now this idea is widely known as ``the induced gravity''.

The thin shell formalism for the generic quadratic gravity was constructed by  H. -H. von Borzeszkowski and V. P. Frolov in Ref.~\refcite{Frolov}. The main difference from General Relativity is that the equations of motion appeared to be the fourth order in derivatives of the metric tensor (this feature is shared by the so called $f(R)$ theories of gravity, and the exceptions are the Horndenski's and Gauss-Bonnet models). This means that the Riemann curvature tensor may have, at most, the jumps across the thin shell (the singular hypersurface $\Sigma$). Therefore, the corresponding extrinsic curvature tensor, $K_{ij}$, must be continuous, i.\,e., 
$[K_{ij}]|_{\Sigma}=0$. But it is the jump of this very tensor, that governs the trajectory of the thin shells in the General Relativity! Thus the matching conditions should also be very different from that ones in General Relativity. And, at least, some of thin shells considered  in the framework of General Relativity, will appear forbidden.

It is very interesting that one particular combination of the quadratic terms is exceptional. This is just the case of the Weyl conformal gravity, with possible addition of the term, linear in curvature scalar, and the cosmological term, i.\,e., the Weyl$+$Einstein gravity, when the quadratic part of the gravitational Lagrangian is just the the square of the Weyl tensor. The matter is that the Weyl tensor itself (with one upper index) is invariant under the local conformal transformation (when the whole metric tensor is multiplied by some function called ``the conformal factor''). The resulting Bach tensor, which appears in the gravitational field equations, is multiplied then by some power of this conformal factor, without any of its derivatives. But, we are free to choose as a conformal factor, anyone of the components of the metric tensor. Thus, we should only require its continuity on the shell, but not the continuity of the derivatives. So, in some cases, the corresponding components of the extrinsic curvature tensor should not be continuous on the shell, and the Riemann curvature tensor constructed from the initial metric (prior to the conformal transformation) may have the delta-function behavior on the thin shell. 

The above consideration is especially important in the case of the spherical symmetry because we are able to use the radius of the sphere (more exactly, its square) as the conformal factor (see Refs.~\refcite{BDE16,BDE16b,bde17a,BDE17,BDE18} for more details), and it is this radius as a function of the proper time on the shell, that determines completely its trajectory in the space-time. 

Recently J. M. M. Senovilla in Ref.~\refcite{Senovilla15} (see also Ref. \refcite{Senovilla17}) discovered that, in the framework of the theory of distributions, the quadratic gravity (as well as all other modifications of the Einstein gravity leading to the fourth order differential equations) allows not only the thin shells (the delta-function distributions), but also the double layers (the delta-prime distributions). This is a very compelling possibility, and, in the absence of the corresponding counterpart in the energy-momentum tensor for the conventional matter (no particles with negative masses), this means that such double layers represent purely geometric phenomenon and describe something like gravitational shock waves (even in the spherically symmetric case, due to the new scalar degree of freedom, what is absolutely forbidden in General Relativity).

The paper is organized as follows. In the Section~2 we introduced the notations and wrote down all the needed formulas and relations. Section~3 devoted to the thorough considerations of the spherical symmetric double layers and the shells in Weyl$+$Einstein gravity. We considered both time-like and null cases (Subsection~3.1). In the Conclusion we summarized the obtained results.

\section{Mathematical introduction}
\label{prelim}

In this Section we describe some needed formalism. We will be using  for the line element the $(+,-,-,-)$--signature
\begin{equation}
ds^2=g_{\mu\nu}dx^\mu dx^\nu \quad (\mu,\nu=0,1,2,3).
\end{equation}
The corresponding metric connections 
\begin{equation}
\Gamma_{\mu\nu}^\lambda=\frac{1}{2} g^{\lambda\sigma}(g_{\sigma\mu,\nu}+g_{\sigma\nu,\mu}-
g_{\mu\nu,\sigma}),
\end{equation}
where comma denotes the ordinary partial derivative. We use the following definition for the the Riemann curvature tensor 
\begin{equation}
R^\mu_{\nu\lambda\sigma}=\frac{\partial\Gamma^\mu_{\nu\sigma}}{\partial x^\lambda}
-\frac{\partial\Gamma^\mu_{\nu\lambda}}{\partial x^\sigma}
+\Gamma^\mu_{\varkappa\lambda}\Gamma^\varkappa_{\nu\sigma}
-\Gamma^\mu_{\varkappa\sigma}\Gamma^\varkappa_{\nu\lambda},
\end{equation}
Respectively, the Ricci tensor equals
\begin{equation}
R_{\mu\nu}=g^{\lambda\sigma}R_{\lambda\mu\sigma\nu}=R^\lambda_{\mu\lambda\nu},
\end{equation}
and the curvature scalar is
\begin{equation}
R=g^{\lambda\sigma}R_{\lambda\sigma}.
\end{equation}
The Einstein tensor
\begin{equation}
G_{\mu\nu}=R_{\mu\nu}-\frac{1}{2}g_{\mu\nu}R.
\end{equation}
The Weyl tensor (which is a completely traceless part of the Riemann tensor) 
\begin{eqnarray}
C_{\mu\nu\lambda\sigma}&=&R_{\mu\nu\lambda\sigma}+\frac{1}{2}(R_{\mu\sigma}g_{\nu\lambda}
+R_{\nu\lambda}g_{\mu\sigma}-R_{\mu\lambda}g_{\nu\sigma}-R_{\nu\sigma}g_{\mu\lambda}) \nonumber \\
&&+\frac{1}{6}R(g_{\mu\nu}g_{\lambda\sigma}-g_{\mu\sigma}g_{\nu\lambda}),  
\end{eqnarray}
The Bach tensor is
\begin{equation}
B_{\mu\nu}=C_{\mu\lambda\nu\sigma}^{\phantom{0000};\sigma;\lambda}
+\frac{1}{2}R^{\lambda\sigma}C_{\mu\lambda\nu\sigma}, 
\end{equation}
\begin{equation}
B^{\lambda}_{\lambda}=0, \quad B_{\mu\nu}=B_{\nu\mu}, \quad B^\lambda_{\mu;\lambda}=0,
\end{equation}
where the semicolon ``;'' denotes the covariant derivative with respect to the metric $g_{\mu\nu}$).

Now we introduce the local conformal transformation by
\begin{equation}
ds^2=\Omega^2(x) d\hat s^2, \quad g_{\mu\nu}=\Omega^2(x) \hat g_{\mu\nu}, \quad 
g^{\mu\nu}=\frac{1}{\Omega^2(x)} \hat g^{\mu\nu},
\end{equation}
where the hat ``$\;\hat{}\;$'' means ``conformally transformed''.

Then, the Einstein tensor is conformally transformed as
\begin{equation}
G_{\mu\nu}=\hat G_{\mu\nu}-\frac{2\Omega_{\mu|\nu}}{\Omega}
+\frac{2\Omega^\lambda_{\phantom{0}|\lambda}}{\Omega}\hat g_{\mu\nu}
+\frac{4\Omega_{\mu}\Omega_{\nu}}{\Omega^2}-\frac{\Omega^\lambda \Omega_\lambda}{\Omega^2}\hat g_{\mu\nu},
\end{equation}
where $\Omega_\mu=\Omega_{,\mu}$, $\Omega^\lambda=\hat g^{\lambda\sigma}\Omega_\sigma$, and the vertical line ``$|$'' means the covariant derivative with respect to the transformed metric $\hat g_{\mu\nu}$. By definition, the Weyl tensor with only one upper index is invariant under the local conformal transformation 
\begin{equation}
C^\mu_{\nu\lambda\sigma}=\hat C^\mu_{\nu\lambda\sigma},
\end{equation}
and so 
\begin{equation}
C^2\sqrt{-g}=C_{\mu\nu\lambda\sigma}C^{\mu\nu\lambda\sigma}\sqrt{-g}=\hat C^2\sqrt{-\hat g},
\end{equation}
Respectively, the Bach tensor is transformed as
\begin{equation}
B_{\mu\nu}=\frac{1}{\Omega^2} \hat B_{\mu\nu}.
\end{equation}
In the decomposition of the total action integral into the gravitational and non-gravitational parts, $S_{\rm tot}=S_{\rm gr}+S_{\rm m}$, the energy -momentum tensor $T_{\mu\nu}$ and its ``transformed'' counterpart
$\hat T_{\mu\nu}$ are both defined through the following variations of $S_{\rm m}$:
\begin{eqnarray}
\delta S_{\rm m}&& \stackrel{def}{=}
\frac{1}{2}\int\! T_{\mu\nu}\sqrt{-g}\delta g^{\mu\nu}dx,  \\
\delta S_{\rm m}&& \stackrel{def}{=} \frac{1}{2}\int\! \hat T_{\mu\nu}\sqrt{-\hat g}\delta \hat g^{\mu\nu}dx,
\end{eqnarray}
Thus, 
\begin{equation}
\hat T_{\mu\nu}=\Omega^2T_{\mu\nu}, \quad \hat T_\mu^\nu=\Omega^4T_\mu^\nu, \quad \hat T^{\mu\nu}=\Omega^6T^{\mu\nu}.
\end{equation}

At this stage, let us take into account the spherical symmetry. The spherical symmetric line element has the form
\begin{equation}
ds^2=g_{\mu\nu}dx^\mu dx^\nu = \gamma_{ik}dx^i dx^k - r^2(x)(d\theta^2+\sin^2\theta d\phi^2),
\quad (i,k=0,1),
\label{ds2}
\end{equation}
where $\theta$ and $\phi$ are the spherical angles, $r(x)$ is the radius of the sphere (with the area $4\pi r^2$), and $\gamma_{ik}$ is the  two-dimensional space-time metric tensor. It appears useful to make the conformal transformation and choose the radius $r(x)$ as the conformal factor, $\Omega(x)=r(x)$. Then,
\begin{equation}
d\hat s^2 = \tilde\gamma_{ik}dx^idx^k - (d\theta^2+\sin^2\theta d\phi^2).
\label{dhats2}
\end{equation}
We make the $(2+2)$--decomposition and, actually, reduce the described problem to the two-dimensional one. Namely,
\begin{eqnarray}
G_{ik}&=&-\frac{2r_{i||k}}{r}+\frac{4r_ir_k}{r^2} \label{Gik}
+\left(1+\frac{2r^p_{\phantom{0}||p}}{r} -\frac{r^p r_p}{r^2}\right)\tilde\gamma_{ik},  \\
G&=&G^\lambda_\lambda=-R=-\frac{1}{r^2}\left(-\hat R+\frac{6r^p_{\phantom{0}||p}}{r}\right), \quad
\hat R=\tilde R-2,
\label{G}
\end{eqnarray}
where $\tilde R$ is the two-dimensional curvature scalar, constructed of the metric tensor $\tilde\gamma_{ik}$, and the double vertical line ``$||$''  denotes the covariant derivative with respect to this very metric.

The corresponding transformed Bach tensor becomes 
\begin{equation}
B_{\mu\nu}=\frac{1}{r^2}\hat B_{ik}, \quad 
\hat B_{ik}=-\frac{1}{6}\left(\tilde R^{||p}_{||p}\tilde\gamma_{ik}-\tilde R_{||ik}+  
\frac{\tilde R^2-4}{4}\tilde\gamma_{ik}\right).
\end{equation}
We do not need expressions for the remaining components, since due to the spherical symmetry $\hat B_2^2=\hat B_3^3$, and the tracelessness of the Bach tensor, $B_\mu^\mu=0$, yields
\begin{equation}
B_2^2=\hat B_3^3=-\frac{1}{2}\hat B_p^p.
\end{equation}

Coming to this point, we should describe the spherically symmetric thin shell, $\Sigma$. It is now a sphere which radius, $\rho$, evolving in time. The corresponding line element equals 
\begin{eqnarray}
ds^2|_\Sigma&=&d\tau^2-\rho^2(\tau) (d\theta^2+\sin^2\theta d\phi^2) \nonumber  \\
&=&\rho^2(\tau)\left(d\tilde\tau^2- (d\theta^2+\sin^2\theta d\phi^2)\right).
\end{eqnarray}
Hence, $\tau$ is the proper time of the observer sitting on this shell (we will call it in the following``the genuine proper time''), while $\tilde\tau$ is its conformal counterpart (we will call it ``the conformal proper time''). The simplest way to proceed further is to use the two-dimensional part of the conformally transformed metric,  
$d\hat s^2 = \tilde\gamma_{ik}dx^idx^k$, $(i,k=0,1)$. Undoubtedly, it is always possible to introduce the Gauss normal coordinate system, associated with the given thin shell. The corresponding line element has the following form, valid both on the shell, as well as just outside it,
\begin{equation}
d\hat s_2^2=\tilde\gamma_{00}(n,\tilde\tau)d\tilde\tau^2-d\tilde n^2,
\label{dhats22}
\end{equation}
where the normal coordinate $n$ runs from inward $n<0$ to outward $n>0$, and the resulting equation of our thin shell is simply $n=0$. To match the above line element with that on the shell, we use the normalization conditions 
\begin{equation}
\tilde\gamma_{00}(0,\tilde\tau)=1.
\end{equation}

The extrinsic curvature tensor, describing the embedding of thin shell (in our case it is just the worldline $\tilde n=0$) into the ambient two-dimensional space-time, is
\begin{equation}
\hat K_{ij}= -\frac{1}{2}\frac{\partial\hat g_{ij}}{\partial n}.
\end{equation}
Now $\hat K_{ij}$ consists of only one component
\begin{equation}
\hat K_{00}\stackrel{def}{=} -\frac{1}{2}\frac{\partial\tilde\gamma_{00}}{\partial n}=\tilde K_{00},
\quad \tilde K=\tilde K^0_0= -\frac{1}{2}\frac{\partial\log\tilde\gamma_{00}}{\partial n},
\end{equation}
and the two-dimensional curvature scalar equals
\begin{equation}
\tilde R=-2\tilde K_n+2\tilde K^2.
\end{equation}

In the Gauss normal coordinates the transformed energy-momentum tensor can be written as 
\begin{equation}
\hat T_{\mu\nu} \stackrel{def}{=} \hat S_{\mu\nu}\delta(n)
+[\hat T_{\mu\nu}]\Theta(n)+\hat T_{\mu\nu}^{(-)},
\end{equation}
where $\hat S_{\mu\nu}$ is the surface energy-momentum tensor on the shell, $\delta(n)$ is the Dirac's delta--function, $\Theta(n)$ is the Heaviside step--function:
\begin{equation}
\Theta(n)=\left\{
\begin{array}{rl}	
1, & \mbox{if } n>0 \;\; (+), \\
0, & \mbox{if } n<0 \;\; (-). 
\end{array}	
\right. 
\end{equation}
We will use the following properties of $\Theta$--function:
\begin{equation}
\Theta^2=\Theta, \quad \Theta'(n)=\delta(n),
\end{equation}
where $[\ldots]=$ denotes the ``jump'' across the shell in the outward normal  direction. Thus,
\begin{equation}
[\hat T_{\mu\nu}]=[\hat T_{\mu\nu}^{(+)}-\hat T_{\mu\nu}^{(-)}].
\end{equation}

\section{Double layers}

The phenomenon of double layer is well known in the classical electrodynamics. It appears naturally in the description of the electric field when there is a probe surface with the opposite charge distributions on its two sides. Formally speaking, in the field equations the double layer manifests itself by appearance of  the derivative of the Dirac $\delta$-function. We have already seen, that such a construction is impossible in General Relativity (Einstein Gravity). Bu it becomes possible in some of its modifications --- the higher derivative gravities. Of course, there exists no counterpart of the electromagnetic dipole in the matter energy-density tensor. Therefore, the gravitational double layer is a pure geometrical phenomenon and can be interpreted as the gravitational shock wave. The main reason why we have chosen just the Weyl$+$Einstein gravity, is the following. We are interested in the description of quantum processes of particle creation caused by gravitational field and its back reaction on the global space-time. Such a back reaction becomes essential in the early universe, near the black hole horizon and the singularity inside and, obviously, on the singular hypersurfaces, i.\,e., thin shells and double layers. To fulfill this program, we constructed the hydrodynamical model allowing the particle creation \cite{BDE16,BDE16b}. Our model differs only slightly from the conventional hydrodynamics written using the Eulerian variables \cite{Ray} in that the constraint describing the particle number conservation is replaced by by the constraint describing the particle creation. Namely, the corresponding part of hydrodynamical action integral undergoes the following surgery
\begin{equation}
\int\lambda_1(nu^\mu)_{;\mu}\sqrt{-g}dx \quad
\longrightarrow \quad \int\lambda_1\left((nu^\mu)_{;\mu}-\beta C^2\right)\sqrt{-g}dx,
\end{equation}
where $\lambda_1$ is the Lagrange multiplier, $(nu^\mu)_{;\mu}$ is the particle production rate --- the number of particle created per unit time per unit volume ($n$ is the invariant particle number density, $u^\mu$ is the four velocity of particle flow), $\beta=const$, and $C^2$ is the square of the Weyl tensor. This creation law was found by Ya. B. Zel’dovich and A. A. Starobinskii in 1977 \cite{Zeld77}. It is remarkable that if one makes the transformation $\lambda_1\to\lambda_1=\gamma_0+\beta\tilde\lambda_1$ with $\gamma_0=const$, then the term $\gamma_0(nu^\mu)_{;\mu}\sqrt{-g}=\gamma_0(nu^\mu)_{,\mu}$ is the full derivative that does not influence the equation of motion, but there emerges the Weyl conformal gravity action integral. For this very reason we will consider, in what follows, the modified gravitational action of the form
\begin{equation}
S_{\rm grav}=-\int\beta\lambda_1C^2\sqrt{-g}dx-\frac{1}{16\pi G}\int(R-2\Lambda)\sqrt{-g}dx.
\end{equation}
Reasonably enough, to obtain the field equations, we need only to replace the Bach tensor $B_{\mu\nu}$, defined above by the modified Bach tensor 
\begin{equation}
B_{\mu\nu}[\lambda_1] = (\lambda_1 C_{\mu \sigma \nu \lambda})^{; \lambda ; \sigma} + \frac{1}{2} \lambda_1 C_{\mu \lambda \nu \sigma} R^{\lambda \sigma},
\end{equation}
the result is
\begin{equation}
4\beta B_{\mu\nu}[\lambda_1]+\frac{1}{16\pi G}G_{\mu\nu}=\frac{1}{2}T_{\mu\nu},
\end{equation}
where $T_{\mu\nu}$ contains only contributions from the matter fields.

After making the conformal transformation with the conformal factor $\Omega=2^2$ and performing the $(2=2)$ decomposition, we are left, actually, with the tw-dimensional metric $\hat ds^2_2=\tilde \gamma_{ik}dx^idx^k$ and the following set of equations
\begin{eqnarray}
\frac{4}{3}\beta\left\{(\lambda_1(\tilde{R}-2))_{||p}^{||p}\tilde{\gamma}_{ik}-(\lambda_1(\tilde{R}-2))_{||ik}+\lambda_1\frac{\tilde{R^2}-4}{4}\tilde{\gamma}_{ik}\right\} && \nonumber  \\
+\frac{1}{8\pi G}\left\{r^2(1+\Lambda r^2)\tilde{\gamma}_{ik}-2r(r_{||ik}-r_{||p}^{||p}\tilde{\gamma}_{ik})+4r_ir_k-r_pr^p\tilde{\gamma}_{ik}\right\} &=&\tilde{T}_{ik} \\ 
2-\tilde{R}+\frac{6r_{||p}^{||p}}{r}=4\Lambda r^2+\frac{8\pi G}{r^2}(\hat{T}^p_p+2\hat{T}^2_2), \quad i,k,p=0,1. &&  
\end{eqnarray}
Here $r_i=r_{,i}$, $r^p=\tilde{\gamma}^{pi}r_i$. , and thje double vertical line ``$||$'' denotes the covariant derivative with respect to metric $\tilde{\gamma}_{ik}$.

Again, let $Sigma$ be a three dimensional hypersurface, where the matter energy-tensor jumps ($=$ shock wave or ``dust-vacuum'' transition) or where some  matter fields are concentrated ($=$ thin shell). Then, one needs to match solutions on both sides of this hypersurface. We will assume now hat $\Sigma$ is time-like (the transition to the space-like case is very easy, and the case of the null hypersurface will be considered separately.)

As before, it is convenient to introduce the Gauss normal coordinates, where the equation for $\Sigma$ is simply $n=0$, and $\hat ds^2_2=\tilde \gamma_{00}(\tau,u)d\tilde\tau^2=du^2$ with $\gamma_{00}(\tau,0)=1$. The transition to the space-like $\Sigma$ is made by interchange ($\tilde\tau\leftrightarrow iu$, $i$ is the imaginary unit). Note, that $\tilde \gamma_{00}$ is continuous across $\Sigma$. 

In this coordinate system the field equations become
\begin{eqnarray}
&&\frac{4}{3}\beta\left\{-(\lambda_1(\tilde{R}-2))_{||nn}\tilde{\gamma}_{00}+\lambda_1\frac{\tilde{R}^2-4}{4}\right\} \nonumber \\
&&+\frac{1}{8\pi G}\left\{r^2(1-\Lambda r^2)\tilde{\gamma}_{00}+3\dot r^2+r_n^2\tilde{\gamma}_{00}\right\}=\hat{T}_{00}  \\
&&-\frac{4}{3}\beta(\lambda_1(\tilde{R}-2))_{||0n}+\frac{1}{8\pi G}\left\{-2rr_{||0n}+4\dot rr_n\right\}=\hat{T}_{0n}  \\ 
&&-\frac{4}{3}\beta\left\{(\lambda_1(\tilde{R}-2))_{||00}\tilde{\gamma}^{00}+\lambda_1\frac{\tilde{R}^2-4}{4}\right\}  \nonumber \\ &&-\frac{1}{8\pi G}\left\{r^2(1-\Lambda r^2)+2r\tilde{\gamma}^{00}r_{||00}-\tilde{\gamma}^{00}\dot r^2-3r_n^2\right\}=\hat{T}_{nn}, \\   
&&2-\tilde{R}+\frac{6}{r}(\tilde{\gamma}^{00}r_{||00}-r_{||nn})=4\Lambda r^2+\frac{8\pi G}{r^2}(\hat{T}^0_0+\hat{T}^n_n+2\hat{T}^2_2).
\end{eqnarray}
Respectively, the trace equation becomes
In these equations $\dot r=r_{,0}$, $ r_n=r_{,n}$. Introducing the the extrinsic curvature tensor $\tilde K_{ij}=-(1/2)\tilde{\gamma}_{ij,n}$, which in our case has only one component, $\tilde K_{00}=-\frac{1}{2}\frac{\partial\tilde{\gamma}_{00,n}}{\partial n}$, we get for the second covariant derivative of any scalar field $\varphi$,
\begin{eqnarray}
\varphi_{||00}&=&\ddot\varphi-\frac{1}{2}\frac{\dot{\tilde\gamma}_{00}}{\tilde{\gamma}_{00}}\dot\varphi+\tilde{\gamma}_{00}\tilde K \varphi_{,n} \\
\varphi_{||0n}&=&\dot\varphi_{,n}-\frac{1}{2}\frac{\dot{\tilde\gamma}_{00}}{\tilde{\gamma}_{00}}\dot\varphi+K \dot\varphi, 
\end{eqnarray}
where $K=\tilde\gamma^{ij}K_{ij}=\tilde\gamma^{00}K_{00}$, while for the curvature scalar $\tilde R$ we have
\begin{equation}
\tilde{R}=-2\tilde{K}_{,n}+2\tilde{K}^2.
\end{equation}
Since $\tilde\gamma_{00}$ is continuous across $\Sigma$, the trace of the extrinsic curvature tensor, $\tilde{K}$, may have, at most, a jump at $n=0$, so $R$ may have, at most, the $\delta$-function behavior there.  And this is possible in General Relativity. Indeed, the Einstein equations in the Gauss normal coordinate system can be obtained from the above ones by simply putting $\beta=0$, the trace equation (the last of them) being the same. But, when $\beta\neq0$, the situation is completely different. Because of the explicit appearance of the $\tilde R^2$ term in the equations, the curvature scalar $\tilde R$ may have, at most, a jump across $\Sigma$ (we are working in the framework of the conventional theory of distributions, where the $\delta^2(n)$) is strictly forbidden). This means that $\tilde K$ is continuous at $\Sigma$, i/\,e., $[\tilde K]\big|_\Sigma=0$, the result discovered by J.~M.~M.~Senovilla \cite{Senovilla15}. More precisely, his statement is that all components of the extrinsic curvature tensor $K_{\mu\nu}$ (without ``hats''!) should be continuous, $[K_{\mu\nu}]\big|_\Sigma=0$, including  $[r_{,n}]\big|_\Sigma=0$. But in our case of the Weyl$+$Einstein gravity, the radius $r$ is decoupled due to the conformal properties of Weyl and Bach tensors, so, it is possible to have $[r_{,n}]\big|_\Sigma\neq0$. Note, that the possibility to have divergent $\tilde R$ in General Relativity and impossibility to have such a behavior in 
Weyl$+$Einstein gravity, indicates the absence of the continuous (analytical) limit $\beta=0$. 

Since $\tilde R$ may have, at most, a jump at $n=0$, its first derivative, $\tilde R_{,n}$,  may contain, at most, a $\delta(n)$ term, and its second derivative --- $\delta'(n)$. It is this $\delta$-prime term that describes the double layer. It should be stressed that in the energy-momentum tensor there is no counterpart to such a term (due to the absence of the mass-dipole), so, the double layers of pure geometrical origin and can be viewed as a gravitational shock wave (even in the case of spherical symmetry!). Thus the energy-momentum tensor has the form
\begin{equation}
\hat{T}_{\mu\nu}=\hat{S}_{\mu\nu}\delta(n)+[\hat{T}_{\mu\nu}]\theta(n)+\hat{T}_{\mu\nu}^{(-)}.
\end{equation}
According to this decomposition one can write
\begin{eqnarray}
r_n&=&[r_n]\Theta(n)+r_n^{(-)}, \\
r_{nn}&=&[r_{n}]\delta(n)+[r_{nn}]\Theta(n)+ r_{nn}^{(-)}, \\ 
{[K]\big|_\Sigma}&=&0, \\
K_{n}&=&[\tilde R]\delta(n)+[\tilde R_{n}]\Theta(n)+ \tilde R_{n}^{(-)}, \\ 
\tilde R&=&[\tilde R]\Theta(n)+\tilde R^{(-)}, \\ 
{[\tilde R_n]}&=&[\tilde R]\delta(n)+[\tilde R_{n}]\Theta(n)+\tilde R_{,n}^{(-)}, \\
{[\tilde R_{nn}]}&=&[\tilde R]\delta'(n)+2[\tilde R_{,n}]\delta(n)+[\tilde R_{,nn}]\Theta(n)+\tilde R_{,nn}^{(-)}, 
\end{eqnarray}
What concerns $\lambda_1$, we will consider it, for the sake of simplicity, as a smooth function.

We see, that the term with $\delta'(n)$ appears only in the $(00)$ equation. All others contain only $\delta(n)$ and $\Theta(n)$. Obeying the rules of the theory of distributions, one should proceed as follows. First, multiple both sides of the given equation by arbitrary function, say $f(\tau,n)$ with a compact support in the normal direction (and, of course, nonzero at $n=0$) and, second, integrate across $\Sigma$, according to definitions: $\int f(\tilde\tau,n)\delta(n)dn=f(\tilde\tau,0)$, $\int f(\tilde\tau,n)\delta'(n)dn=-f'(\tilde\tau,0)$. Evidently, the $\delta(n)$-function is concentrated on $\Sigma$, so, in the absence of $\delta'(n)$ one can simply equate the coefficients on both sides of the corresponding equation in front of, separately, $\delta(n)$ and $\Theta(n)$. And now we will do it for $(0n)$ and $(nn)$ and (trace) equations. For the $(0n)$-equation we have, omitting continuous parts, 
\begin{eqnarray}
&&-\frac{4}{3}\beta\left\{[\lambda_1(\tilde{R}-2)]\,\dot{}\, \delta(n)+[\lambda_1(\tilde{R}-2)]\,\dot{}_{\!,n}\,\Theta(n)
+\tilde K[\lambda_1(\tilde{R}-2)]\,\dot{}\,\Theta(n) \right\}  \nonumber \\ 
&&+\frac{1}{4\pi G}\left\{-r[r_n]\,\dot{}\,\Theta(n)+2\dot r[r_n]\Theta(n)\right\}=\hat{S}_{nn}\delta(n)+[\hat{T}_{0n}]\Theta(n),
\end{eqnarray}
and the result is
\begin{eqnarray}
&&-\frac{4}{3}\beta[\lambda_1(\tilde{R}-2)]\,\dot{}=\hat{S}_{0n}, \\
&&-\frac{4}{3}\left\{\beta[\lambda_1(\tilde{R}-2)]\,\dot{}_{\!,n}+\tilde K(\tilde{R}-2)]\,\dot{}\right\}+\frac{1}{4\pi G}\left\{-r[r_n]\,\dot{}+2\dot r[r_n]\right\}=[\hat{T}_{0n}].
\end{eqnarray}
This last equation we will need when deriving the continuity equation for the shell. What concerns the $(nn)$ and trace equations, we are interested only in the $\delta$-function contribution, and get the following result
\begin{eqnarray}
&&-\frac{4}{3}\beta\tilde K[\lambda_1(\tilde{R}-2)]=\hat{S}_{nn}=-\hat{S}_n^n, \\
&&-[r_n]=-\frac{4\pi G}{3r}\left(\hat{S}_0^0+\hat{S}_n^n+2\hat{S}_2^2\right).
\end{eqnarray}

At last, let us consider the remaining $(00)$ equation, the most important for us. Multiplying both sides of it by function $f(\tilde\tau,n)$ and integrating across $\Sigma, we obtain$
\begin{eqnarray}
&&-\frac{4}{3}\beta\int f(\tilde\tau,n)\left\{[\lambda_1(\tilde{R}-2)]\delta'(n)
+2[\lambda_1(\tilde{R}-2)]\delta(n)\right\}dn \nonumber \\
&&-\frac{1}{4\pi G}\int f(\tilde\tau,n)r[r_n]\delta(n)dn=f(\tilde\tau,0)\tilde S_{00}.
\end{eqnarray}
Eventually, we get
\begin{equation}
\frac{4}{3}\beta\left\{-[\lambda_1(\tilde{R}-2)]_{,n}+\varphi(\tilde\tau)\right\}
-\frac{1}{4\pi G}r[r_n]=\tilde S_{00},
\end{equation}
where $\varphi(\tilde\tau)=f'(\tilde\tau,0)/f(\tilde\tau,0)$ is the arbitrary function of the conformal proper time $\tilde\tau$. 
Let us summarize, what we have got till now. The full set of the equations on the hypersurface $\Sigma$ looks as follows:
\begin{equation}
\left\{
\begin{array}{rl}	
&\frac{4}{3}\beta\left\{-[\lambda_1(\tilde{R}-2)]_{,n}+\varphi(\tilde\tau)\right\}
-\frac{1}{4\pi G}r[r_n]=\tilde S_0^0,  \\
&-\frac{4}{3}\beta[\lambda_1(\tilde{R}-2)]\,\dot{}=\hat{S}_{0n}, \\
&\frac{4}{3}\beta\tilde K[\lambda_1(\tilde{R}-2)]=\hat{S}_n^n, \\
&\frac{3}{4\pi G}r[r_n]=\hat{S}_0^0+\hat{S}_n^n+2\hat{S}_2^2.
\end{array}	
\right. \label{four}
\end{equation}
Differentiating the first of these equations in time variable $\tilde\tau$ and using all others plus the equation for $[T_{0n}]$, we come to the energy conservation equation of the form
\begin{equation}
\dot{\hat{S}}^0_0-(\varphi(\tilde\tau)+\tilde K)
-\frac{4}{3}\beta(\dot\varphi(\tilde\tau)
+\frac{\dot r}{r}\tilde K)[\lambda_1(\tilde{R}-2)]
-\frac{\dot r}{r}(\tilde S_0^0+2\tilde S_2^2)+[T_0^n]=0.
\end{equation}
Note, that if $\beta=0$, we would reproduce exactly the thin shell equations of General relativity. Note als, that if $\tilde R$ is continuous across $\Sigma$, but $\tilde R_{,n}$ is not, i.\,e., $[R_{,n}]\neq0$, then the first equation  in (\ref{four}) would become 
\begin{equation}
-\frac{4}{3}\beta[\lambda_1(\tilde{R}-2)]_{,n}-\frac{1}{4\pi G}r[r_n]=\tilde S_0^0,
\end{equation}
while starting from the field equations before the integration across $\Sigma$, we would obtain
\begin{equation}
-\frac{8}{3}\beta[\lambda_1(\tilde{R}-2)]_{,n}-\frac{1}{4\pi G}r[r_n]=\tilde S_0^0.
\end{equation}
Therefore, when sitting on the hypersurface $\Sigma$, it is impossible to distinguish whether we are dealing with the double layer, or just with some contribution from the Weyl tensor to the thin shell without the double layer. To do this, one has to look a little bit outside the shell.

The most striking consequence of the existence of the double layer is the possibility to have nonzero $\hat S^n_n$ and $\hat S^n_0$. This feature was especially noted by J.~M.~M.~Seno\-villa \cite{Senovilla17}, though he found no physically acceptable explanation for this phenomenon. (Remember, that in General Relativity $\hat S^n_n=\hat S^n_0=0$ is required by the matching conditions.) We propose the following solution to this problem. The double layer itself creates particles, some of them leaving the hypersurface $\Sigma$, there energy flow being described by $S^n_0=0$, and the momentum transfer --- by
$S^n_n=0$.

It appeared that there is one more important consequence of appearance of the double layer, i.\,e., $\delta'(n)$-function in the field equations. For given function$\varphi(\tilde\tau)$ in the matching conditions, which is uniquely defined for any particular problem, the result of the integration across $\Sigma$ depends on weather we raise the index ``0'' by metric coefficient $\tilde\gamma^{00}$ before this integration, or after. The additional term is proportional to the trace of extrinsic curvature tensor. To avoid the ``hysteresis'' we have to put it zero. Thus
\begin{equation}
\tilde K\big|_\Sigma=0,
\end{equation}
and this seems the more severe requirement. From it follows immediately, that
\begin{equation}
\hat S_n^n=0.
\end{equation}
Note, that such a result supports our proposal for the interpretation of the $S_0^n$ and $S_n^n$ coefficients of the surface energy-momentum tensor: from the point of view of the observer, sitting on $\Sigma$ (the comoving observer), particles should be produced in pairs having the opposite momenta.

\subsection{Null double layer}

It is already mentioned that it is impossible to introduce the Gauss normal coordinate system in the case of the null hypersurface $\Sigma$, because the normal vector lies in in this very surface. Instead, we will use the the double null coordinates, $(u,v)$. The conformally transformed two-dimensional line element takes now the form 
\begin{equation}
d\tilde s^2_2= \tilde\gamma_{ik}dx^i dx^k =2H(u,v)dudv
\end{equation}
with the only nonzero component of the metric tensor $\tilde\gamma_{uv}=H$, and its inverse is $\tilde\gamma^{uv}=1/H$. Our field equations now look as follows
\begin{eqnarray}
&&-\frac{4}{3}\beta\left\{(\lambda_1(\tilde{R}-2))_{uu}
-\frac{H_{,u}}{H}(\lambda_1(\tilde{R}-2))_{u}\right\} \nonumber \\
&&-\frac{1}{4\pi G}\left\{rr_{uu}-\frac{H_{,u}}{H}rr_{u}-2r_{,u}^2\right\}=\hat T_{uu}, \\
&&-\frac{4}{3}\beta\left\{(\lambda_1(\tilde{R}-2))_{vv} 
-\frac{H_{,v}}{H}(\lambda_1(\tilde{R}-2))_{v}\right\} \nonumber \\
&&-\frac{1}{4\pi G}\left\{rr_{vv}-\frac{H_{,v}}{H}rr_{v}-2r_{,v}^2\right\}=\hat T_{vv}, \\
&&-\frac{4}{3}\beta\left\{(\lambda_1(\tilde{R}-2))_{uv} 
-\frac{\lambda_1}{4}(\tilde{R}-4)H\right\} \nonumber \\
&&-\frac{1}{8\pi G}\left\{r^2(1- \Lambda r^2)H+2(rr_{,uv}+r_ur_v)\right\}=\hat T_{uv}, \\
&&2-\tilde{R} -\frac{12}{H}\frac{r_{,uv}}{r}=
4 \Lambda r^2+\frac{8\pi G}{r^2}\left(\hat T_u^u+\hat T_v^v+\hat T_2^2\right).
\end{eqnarray}
Note, that $T_u^u=(1/H)T_{uv}$, $T_v^v=(1/H)T_{uv}$, so $T_u^u=T_v^v$. The two-dimensional curvature scalar equals now
\begin{equation}
\tilde R=-\frac{2}{H^3}(HH_{,uv}-H_uH_v).
\end{equation}
We see, that $H_{,uu}$ or $H_{,vv}$ does not enter this expression.

For definiteness, we assume that our null hypersurface $\Sigma$ lies at $u=0$, and the null coordinate $v$ runs along it. Then the matter energy-momentum tensor can be written as
\begin{equation}
\hat T_{\mu\nu}=\hat S_{\mu\nu}\delta(u)+[\hat T_{\mu\nu}]\Theta(u)+\hat T_{\mu\nu}^{(-)}.
\end{equation}
The $(vv)$-equation does not contain $\delta(u)$-function at all. It follows immediately that 
\begin{equation}
\hat S_{uv}=0.
\end{equation}
The same is true for the trace equation, So
\begin{equation}
\hat S_u^u=S_v^v=S_u^u=-S_2^2.
\end{equation}
What concerns the $(uv)$-equation, it is possible to have $\delta(u)$-function in its left hand side --- nothing more singular. We get 
\begin{equation}
\frac{4}{3}\beta[(\lambda_1(\tilde{R}-2))_{v}]=\hat S_{uv}.
\end{equation}
Let us turn now to the remaining $(uu)$-equation. It is not difficult to see that, provided the two-dimensional curvature scalar is jumped at $\Sigma$ (and so does $H_{,u}$), then we will encounter the product of two distributions, namely $H_{,u}\Theta(u)\delta(u)$. And this is forbidden by the conventional theory. Thus, we have to require
\begin{equation}
[H_{,u}]=0.
\end{equation}
But! In such a case
\begin{equation}
[\tilde R]=0.
\end{equation}
Since we decided to consider the Lagrangian multiplier $\lambda_1$ to be a smooth enough, this means that the null double layer in the Weyl$+$Einstein gravity simply does not exist!
	
\section{Conclusion}

In this paper we undertook the thorough investigation of the matching conditions on the singular hypersurfaces in Weyl$+$Einstein gravity. 

Unlike in the General Relativity, where the singular hypersurface may contain, at most, the Dirac $\delta$-function both in the matter energy-momentum tensor in the right hand side of the Einstein equations and in the second derivatives of the metric tensor in their left hand side, in the so called quadratic gravity (the gravitational action integral, except the Einstein-Hilbert and cosmological terms, has all possible quadratic combinations of the Riemann curvature scalar). there appear the possibility of the double layer, i.\,e., the derivative of the $\delta-function$.

But, this derivative is absent in the energy-momentum tensor (no mass-dipole analogous to the charge-dipoles in the classical electrodynamics), so the double layer is a purely geometrical phenomenon ant it may be treated as the purely gravitational shock wave.

The mathematical formalism for the double layers was elaborated layers was elaborated by J.~M.~M~Senovilla for the double layers was elaborated for the generic quadratic gravity. Our choice  of the Weyl$+$Einstein gravity is motivated by thee fact that the latter differs from the generic case in some aspects and requires separate consideration. Moreover, we confined ourselves by the spherically symmetry, because in such a case the theory becomes, essentially, two-dimensional. The three-dimensional hypersurface reduces to the world-line, and it becomes much easier to understand every step in the calculations and interpretations of the results.

The main results are the following. We derived the matching conditions for the spherically symmetric singular hypersurface (in our case it is equivalent to the world line) in the Weyl$+$Einstein gravity. It was found, that the residual extrinsic curvature tensor of this surface (i.\,e., having only one, $(00)$)-component after separating the spherical angular part of the metric tensor), indeed, must be continuous on the singular hypersurface. The result is the same as that found by Senovilla, it is dictated bt the very possibility to have the double layer, but the jump in the normal derivative of the radius may be not zero. (and this is different from Senovilla's). 

It was found that in the presence of the double layer, the matching conditions contain an arbitrary function (in our case it is a function of the proper conformal time of the observer, sitting on the singular world-line), and this result is quite new and very important. One of the consequences of such freedom is that the trace of the extrinsic curvature tensor of the singular hypersurface is necessarily equal to zero.

We suggested the physical interpretation for the $S_0^n$ and $S_n^n$ components of the surface matter energy-momentum tensor of the shell. In General Relativity they are zero by virtue of the Einstein equations. In the quadratic gravity they are not necessarily zero. Our interpretation is that these components describe the energy flow $(S^{0n})$ and the momentum transfer $(S^{nn})$ of the particles produced by the double layer itself. Moreover, the requirement of the zero trace of the extrinsic curvature tensor (mentioned above) implies that $S_n^n=0$, and this fact also support our suggestion, because it means that for the observer sitting on the shell, the particles will be seen created by pairs, and the sum of their momentum transfers must zero.

We derived also the matching conditions for the null hypersurface, and this is, again, quite new. We found that the null-double layer in the Weyl$+$ Einstein gravity does not exist at all.

\section*{Acknowledgments}

We acknowledge E. O. Babichev and A. L. Smirnov for helpful discussions. The authors acknowledge support from RFBR 18-52-15001 NCNIa (Russia) and from the research program ``Programme national de cosmologie et galaxies'' of the CNRS/INSU (France).


\begin{thebibliography}{00}   
	
\bibitem{Isr} W. Israel, {\it Il Nuovo Cim. B} {\bf 44} (1966) 1.
	
\bibitem{Cruz67} V. de la Cruz and W.Israel, {\it Il Nuovo Cim.} {\bf 51} (1967) 744.
	
\bibitem{Lake79} K. Lake, {\it Phys. Rev. D} {\bf 19} (1979).
	
\bibitem{BKT83} V. A. Berezin, V. A. Kuzmin and I. I. Tkachev, {\it Phys. Lett. B} {\bf 120} (1983) 91.
	
\bibitem{BKT83b} V. A. Berezin, V. A. Kuzmin and I. I. Tkachev, {\it Phys. Lett. B} {\bf 124} (1983) 479.
	
\bibitem{BKT83c} V. A. Berezin, V. A. Kuzmin and I. I. Tkachev, {\it Phys. Lett. B} {\bf 130} (1983) 23.
	
\bibitem{BKT84} V. A. Berezin, V. A. Kuzmin and I. I. Tkachev, {\it Sov. Phys.—JETP} {\bf 59} (1984) 459.
	
\bibitem{Gyulassy84} M. Gyulassy, K. Kajantie, H. Kurki-Suonio and L. McLerran. {\it Nucl. Phys. B} {\bf 237} (1984) 477. 
	
\bibitem{Lake84} K. Lake, {\it Phys. Rev. D} {\bf 29} 1861 (1984).
	
\bibitem{Grand84} T. De Grand and K.Kajantie, Phys. Lett. 147B, 273 (1984). [5] A.Aurilia, G.Denardo,
	
\bibitem{Legovini84} F. Legovini and E. Spallucci, Phys. Lett. 147B, 258 (1984), Nucl. Phys. B252, 523 (1985).
	
\bibitem{BKT85} V. A. Berezin, V. A. Kuzmin and I. I. Tkachev, {\it Pis’ma Zh. Eksp. Teor. Fiz.} {\bf 41} (1985) 446.
	
\bibitem{Sato86} H. Sato, {\it Progr. Theor. Phys.} {\bf 76} (1986) 1250. 
	
\bibitem{Laguna86} P. Laguna-Castillo and R. A. Matzner, {\it Phys. Rev. D} {\bf 34}, 2913 (1986).
	
\bibitem{BKT87} V. A. Berezin, V. A. Kuzmin and I. I. Tkachev, {\it Sov. Phys. —JETP} {\bf 93} (1987) 1159.
	
\bibitem{BKT87b} V. A. Berezin, V. A. Kuzmin and I. I. Tkachev, {\it Phys. Rev. D} {\bf 36} (1987) 2919.
	
\bibitem{Blau87} S. K. Blau, E. I. Guendelman and A. H. Guth {\it Phys. Rev. D} {\bf 35} (1987) 1747.
	
\bibitem{Farhi87} E. Farhi and A. Guth, Phys. Lett. B183, 149 (1987).
	
\bibitem{Aurilia89} A. Aurilia, M. Palmer and E. Spallucci, {\it Phys. Rev. D} {\bf 40} (1989) 2511.
	
\bibitem{Farhi90} E. Farhi, A. H. Guth and J. Guven, {\it Nucl.Phys. B} {\bf 339} (1990) 417.
	
\bibitem{BKT90}  V. A. Berezin, V. A. Kuzmin and I. I. Tkachev, {\it Int. J. Mod. Phys. A} {\bf 45} (1990) 4639. 
	
\bibitem{BKT91}  V. A. Berezin, V. A. Kuzmin and I. I. Tkachev, {\it Physica Scripta} {\bf 36} (1991) 269.
	
\bibitem{BKT91b} V. A. Berezin, V. A. Kuzmin and I. I. Tkachev, {\it Phys. Rev. D} {\bf 43} (1987) R3112.
	
\bibitem{Aguirre05} A. Aguirre and M. Johnson, {\it Phys. Rev. D} {\bf 72} (2005) 103525.
	
\bibitem{Lee07} B-H. Lee, W. Lee, S. Nam and C. Park, {\it Phys. Rev. D} {\bf 75} (2007) 103506.
	
\bibitem{Mann08} R. B. Mann and J. J. Oh, {\it Phys. Rev. D} {\bf 74} (2006) 124016; Erratum {\it Phys. Rev. D} {\bf 77} (2008) 129902.
	
\bibitem{Boulware73} D. G. Boulware, {\it Phys. Rev. D} {\bf 8} (1973) 2363.
	
\bibitem{Frolov74} V. P. Frolov, {\it Sov. Phys. JETP} {\bf 38} (1974) 393. 
	
\bibitem{Berezin05} V. Berezin, V. Dokuchaev, Y. Eroshenko and A. Smirnov {\it Class. Quant. Grav.} {\bf 22} (2005) 4443.
	
\bibitem{Dokuch07} V. I. Dokuchaev  and S. V. Chernov 2007 {\it J. Exp. Theor. Phys. Lett.} {\bf 85} (2007) 595.
	
\bibitem{Dokuch08} V. I. Dokuchaev  and S. V. Chernov  {\it J. Exp. Theor. Phys.} {\bf 107} (2008) 203.
	
\bibitem{Chernov08} S. V. Chernov and V. I. Dokuchaev {\it Class. Quant. Grav.} {\bf 25} (2008) 015004.
	
\bibitem{Dokuch10} V. I. Dokuchaev and S. V. Chernov,  {\it J. Exp. Theor. Phys.} {\bf 111} (2010) 570.
	
\bibitem{Berezin14} V. A. Berezin, V. I. Dokuchaev, arXiv:1404.2726 [gr-qc].
	
\bibitem{Berezin14b} V. A. Berezin, V. I. Dokuchaev, arXiv:1404.2727 [gr-qc].
	
\bibitem{FMM} V. P. Frolov, M. A. Markov  and V. F. Mukhanov, {\it Phys. Rev. D} {\bf 41} (1990) 383.

\bibitem{Berezin} V. A. Berezin, in {\sl General Relativity and Gravitation, Vol. 1, Classical Relativity. Proc. 10th Int. Conf. on General Relativity and Gravitation, Padova, Italy} eds. B. Bertotti, F. de Felice, A. Pascolini (Consiglio Nazionale delle Ricerche, Rome, 1983) p. 685.
	
\bibitem{BI} C. Barrabes and W. Israel, {\it Phys. Rev. D} {\bf 43} (1991) 1129. 

\bibitem{Parker69} L. Parker, {\it Phys. Rev.} {\bf 183} (1969) 1057.
	
\bibitem{Grib70}  A. A. Grib and S. G. Mamaev, {\it Sov. J. Nucl. Phys.} {\bf 10} (1970) 722.
	
\bibitem{Zeld70} Ya. B. Zel’dovich, {\it JETP Lett.} {\bf 9} (1970) 307.
	
\bibitem{Zeld71} Ya. B. Zel’dovich and L. P. Pitaevsky, {\it Comm. Math. Phys.} {\bf 23} (1971) 185.
	
\bibitem{Zeld72} Ya. B. Zel’dovich and A. A. Starobinskii, {\it Sov. Phys. JETP} {\bf 34} (1972) 1159.
	
\bibitem{Parker73} L. Parker and S. A. Fulling, {\it Phys. Rev. D} {\bf 7} (1973) 2357.
	
\bibitem{Fulling73} S. A. Fulling, {\it Phys. Rev. D} {\bf 7} (1973) 2850.
	
\bibitem{Parker73b} B. L. Hu, S. A. Fulling and L. Parker, {\it Phys. Rev. D} {\bf 8} (1973) 2377.
	
\bibitem{Parker74} L. Parker and S. A. Fulling, {\it Phys. Rev. D} {\bf 9} (1974) 341.
	
\bibitem{Parker74b} S. A. Fulling, L. Parker and B. L. Hu, {\it Phys. Rev. D} {\bf 10} (1974) 3905.
	
\bibitem{Parker74c} S. A. Fulling and L. Parker, {\it Ann. Phys.} {\bf 87} (1974) 176.
	
\bibitem{Lukash74} V. N. Lukash and A. A. Starobinskii, {\it Sov. Phys. JETP} {\bf 39} (1974) 742.
	
\bibitem{Zeld77}  Ya. B. Zel’dovich and A. A. Starobinskii, {\it JETP Lett.} {\bf 26} (1977) 252.
	
\bibitem{Alstar} A. A. Starobinsky, {\it Phys. Lett. B}{\bf 91} (1980) 99.
	
\bibitem{Sakh} A. D. Sakharov, {\it Sov. phys. Doklady} {\bf 12} (1968) 1040 [in Russian: DAN (1967)].
	
\bibitem{Frolov} H. -H. von Borzeszkowski and V. P. Frolov, {\it Annalen der Physik} {\bf 37} (1980) 285.
	
\bibitem{BDE16} V. Berezin, V. Dokuchaev and Yu. Eroshenko, {\it JCAP} {\bf 01} (2016) 019.
	
\bibitem{BDE16b} V. Berezin, V. Dokuchaev and Yu. Eroshenko, {\it J. Mod. Phys. A} {\bf 31} (2016) 1641004.
	
\bibitem{bde17a}  V. A. Berezin, V. I. Dokuchaev, Yu. N. Eroshenko, Russian Phys. J. {\bf 59} (2017) 1819.
	
\bibitem{BDE17} V. Berezin, V. Dokuchaev and Yu. Eroshenko, {\it JCAP} {\bf 01} (2017) 018.

\bibitem{BDE18} V. Berezin, V. Dokuchaev and Yu. Eroshenko, {\it Int. J. Mod. Phys. D} {\bf 27} (2018) 1841012.

	
\bibitem{Senovilla15} J. M. M. Senovilla, {\it J. of Phys. Conf. Ser.} {\bf 600} (2015) 012004.
	
\bibitem{Senovilla17} E. F. Eiroa, G. F. Aguirre and J. M. M. Senovilla, {\it Phys. Rev. D} {\bf 95} (2017) 124021.

	
\end{thebibliography}
\end{document}